\documentclass[12pt,letterpaper]{JHEP3}

\usepackage{amsmath,amssymb,mathrsfs}

\def\linebreak{\hfill\break}

\def\bra<#1|{\langle #1\rvert}
\def\ket|#1>{\lvert#1 \rangle}
\def\braket<#1|#2>{\langle #1|#2 \rangle}

\def\pfrac#1#2{\left(\frac{#1}{#2}\right)}
\def\const{\text{const}}
\def\otop#1{\hbox{$#1\kern-0.1em$\llap{\hbox{\raise1.7ex\hbox{$\scriptstyle\circ$}}}} }

\def\inpare#1{\left(#1\right)}
\def\bigpare(#1){\left(#1\right)}

\def\insbra#1{\left[ #1 \right]}
\def\bigbra[#1]{\left[ #1 \right]}

\def\therefore{\mbox{\setbox0=\hbox{X}\hbox{$\ldotp$}\raise0.7\ht0\hbox{$\ldotp$}\hbox{$\ldotp$}} \quad }
\def\because{\mbox{\setbox0=\hbox{X}\raise0.7\ht0\hbox{$\ldotp$}\hbox{$\ldotp$}\raise0.7\ht0\hbox{$\ldotp$}}\kern0pt }

\def\r#1{{\rm #1}}

\def\ZR{{{\mathbb Z}}}

\def\maps{\rightarrow}

\def\Frac(#1/#2){\left(\frac{#1}{#2}\right)}

\def\Im{{\rm Im\,}}
\def\Re{{\rm Re\,}}

\def\w{\wedge}

\def\pd{\partial}

\def\dual{{}*\! }

\def\Eq#1{\begin{equation} #1 \end{equation}}

\def\Eqr#1{\begin{eqnarray} #1 \end{eqnarray}}

\def\Eqrsub#1{\begin{subequations}\Eqr{#1}\end{subequations}}
\def\Eqrsubl#1#2{\begin{subequations}\label{#1}\Eqr{#2}\end{subequations}}
\def\Bitm{\begin{itemize}}
\def\Eitm{\end{itemize}}
\def\Blist#1#2{\begin{list}{#1}{\parsep=0pt \itemsep=0pt%
  \listparindent=0pt #2}}
\def\Elist{\end{list}}
\long\def\ignore#1#2{\def\ignoreflag{#1}\long\def\tmptext{#2}
  \ifnum\ignoreflag>1 #2 \fi}

\newcommand{\nn}{\nonumber}

\def\Xsp{{\rm X}_4}
\def\bXsp{\bar{{\rm X}}_4}
\def\Ysp{{\rm Y}_6}

\def\tXsp{\tilde{{\rm X}}_{10}}

\numberwithin{equation}{section}

\title{Comments on the four-dimensional effective theory 
for warped compactification}

\author{
Hideo Kodama and Kunihito Uzawa\\ 
Yukawa Institute for Theoretical Physics\\
Kyoto University, Kyoto 606-8502, Japan.\\
E-mail: \email{kodama@yukawa.kyoto-u.ac.jp, 
uzawa@yukawa.kyoto-u.ac.jp }}

\abstract{
We derive four-dimensional effective theories for warped 
compactification of the ten-dimensional IIB supergravity 
and the eleven-dimensional Ho\v{r}ava-Witten model. 
We show that these effective theories allow a much wider 
class of solutions than the original higher-dimensional 
theories. In particular, the effective theories have 
cosmological solutions in which the size of the internal 
space decreases with the cosmic expansion in the Einstein 
frame. This type of compactifying solutions are not allowed 
in the original higher-dimensional theories. This result 
indicates that the effective four-dimensional theories 
should be used with caution, if one regards the higher-dimensional theories more fundamental.
} 
\preprint{YITP-05-73 \hfill\\{\tt hep-th/0512104}}

\begin{document}
%
%
\section{Introduction}

Recently, a new class of dynamical solutions describing a 
size-modulus instability in the ten-dimensional type IIB 
supergravity model have been discovered by Gibbons et al. 
\cite{Gibbons:2005rt} and the authors \cite{Kodama:2005fz}. These 
solutions can be always obtained by replacing the 
constant modulus $h_0$ in the warp factor 
$h=h_0+h_1(y)$ for supersymmetric solutions by a linear function 
$h_0(x)$ of the four-dimensional coordinates $x^\mu$.
Such extensions exist for many of the well-known solutions 
compactified with flux on a conifold, resolved conifold, deformed 
conifold and compact Calabi-Yau manifold  \cite{Kodama:2005fz}.

In most of the literature, the dynamics of the internal space, 
namely the moduli, in a higher-dimensional theory is investigated by 
utilising a four-dimensional effective theory. In particular, 
effective four-dimensional theories are used in essential ways in 
recent important work on the moduli stabilisation problem and the 
cosmological constant/inflation problem in the IIB sugra 
framework \cite{Kachru:2003aw,Kachru:2003sx,Klebanov:2000hb,
Witten:1996bn,Kachru:2000ih,Tripathy:2002qw}. 
Hence, it is desirable to find the relation between the
 above dynamical solutions in the higher-dimensional 
theories and solutions in the effective four-dimensional theory. 

In the conventional approach where the non-trivial warp factor does 
not exist or is neglected, an effective four-dimensional theory is 
derived from the original theory  assuming the ``product-type'' 
ansatz for field variables \cite{Salam:1981xd,Duff:1986hr}. This 
ansatz requires that each basic field of the theory is expressed as 
the sum of terms of the form $f(x)\omega(y)$, where $f(x)$ is an 
unknown function of the four-dimensional coordinates $x^\mu$, and 
$\omega(y)$ is a known harmonic tensor on the internal space. 
Further, it is assumed that the higher-dimensional metric takes the 
form $ds^2=ds^2(\Xsp)+ h_0^{\beta}(x)ds^2({\rm Y})$, where 
$ds^2(\Xsp)=g_{\mu\nu}(x)dx^\mu dx^\nu $ is an unknown four-dimensional 
metric, $h_0(x)$ is the size modulus for the internal space 
depending only on the $x$-coordinates, and $ds^2({\rm Y})=\gamma_{pq}dy^p 
dy^q$ is a (Calabi-Yau) metric of the internal space that depends on 
the $x$-coordinates only through moduli parameters. Under this 
ansatz, the four-dimensional effective action is obtained by 
integrating out the known dependence on $y^p$ in the 
higher-dimensional action. 

The dynamical solutions in the warped compactification mentioned at 
the beginning, however, do not satisfy this ansatz. Hence, in order 
to incorporate such solutions to the effective theory, we have to 
modify the ansatz. Taking account of the structure of the 
supersymmetric solution, the most natural modification of the ansatz 
is to introduce the non-trivial warp factor $h$ into the metric as 
$ds^2=h^{\alpha} ds^2(\Xsp)+ h^{\beta} ds^2({\rm Y})$ and assume that $h$ 
depends on the four-dimensional coordinates $x^\mu$ only through the 
modulus parameter of the supersymmetric solution as in the case of 
the internal moduli degrees of freedom. This leads to the form 
$h=h_0(x)+h_1(y)$ for the IIB models, which is consistent with the 
structure of the dynamical solutions in the ten-dimensional theory.

In the present paper, starting from this modified ansatz, we study 
the dynamics of the four-dimensional effective theory and its 
relation to the original higher-dimensional theory for warped 
compactification of the ten-dimensional type IIB supergravity and 
the eleven-dimensional Ho\v rava-Witten model. For simplicity, we 
assume that the moduli parameters other than the size parameter are 
frozen. 

The paper is organised as follows. First, in the next section, we 
discuss the dynamics of the size modulus and the spacetime for 
compactification with vanishing flux in the ten-dimensional type IIB 
supergravity, starting from the standard "product-type" ansatz, for 
comparison. We show that the four-dimensional effective theory in 
this case is equivalent to the original ten-dimensional theory under 
the ansatz. Then, in the following two sections, we derive the 
four-dimensional effective theory for warped compactifications 
starting from the modified ansatz and compare it with the original 
higher-dimensional theory. The compactification on a compact 
Calabi-Yau manifold in the ten-dimensional IIB supergravity is 
treated in \S\ref{sec:flux compactification},  
and the Ho\v rava-Witten model of the eleven-dimensional heterotic 
M-theory is discussed in \S\ref{sec:Horava-Witten}. In both of these 
models, it is shown that the four-dimensional effective theory 
contains spurious solutions that are not allowed in the original 
higher-dimensional theory. Finally, Section \ref{sec:conclusion} is 
devoted to summary and discussion. 
%
%
%
\section{Compactification with vanishing flux in the 10D 
supergravity}
 \label{sec:no-flux compactification}

When all form fluxes vanish and the dilaton is constant, 
the ten-dimensional supergravity reduces to the vacuum 
Einstein equations in ten dimensions, irrespective of 
the type of the theory. In this reduced theory, the 
direct product of the four-dimensional Minkowski spacetime 
and a six-dimensional Calabi-Yau space provides a supersymmetric 
solution. In this section, we briefly discuss the 
four-dimensional effective theory for this simple 
compactification, for comparison with the cases of flux 
compactification studied in the subsequent sections

%
%
\subsection{Ansatz and a general solution}
  \label{subsec:noflux-ansatz}

Let us consider the ten-dimensional spacetime with the metric
%
%
\begin{eqnarray}
ds^2(\tXsp)&=&h_0^{-1/2}(x)\,ds^2(\Xsp)
       +h_0^{1/2}(x)\,ds^2(\Ysp),
 \label{eq:nw-general-metric}
\end{eqnarray}
%
where $\Xsp$ is the four-dimensional spacetime with 
coordinates $x^\mu$, and $\Ysp$ is the six-dimensional 
internal space. We assume that there exists no flux and 
the dilaton is constant. Then, if $\Xsp$ is flat, $\Ysp$ 
is a Calabi-Yau manifold, and $h_0$ is a constant, this 
metric gives a supersymmetric solution to the ten-dimensional 
supergravity, and $h_0$ can be regarded as the parameter 
representing the size modulus of the internal space. 
Hence, when we discuss the four-dimensional dynamics of this 
size modulus, the metric \eqref{eq:nw-general-metric} provides 
the most natural class, for which $h_0$ depends only on the 
coordinates $x^\mu$ of the four-dimensional spacetime and 
$ds^2(\Ysp)$ is some fixed metric on $\Ysp$ that does not 
depend on $x^\mu$ . 

Since we are assuming that all gauge fields vanish and the 
dilaton is constant, the dynamics is completely determined 
by the ten-dimensional vacuum Einstein equations, which 
read in the present case as
%
%
\Eqrsubl{eq:einstein-nw}{
& &R_{\mu\nu}(\Xsp)-h_0^{-1}\,D_{\mu}D_{\nu}h_0
  +\frac{1}{4}g_{\mu\nu}(\Xsp)h_0^{-1}\triangle_{{\rm X}}h_0=0,
  \label{eq:einstein-nw:a}
\\
& &R_{pq}(\Ysp)-\frac{1}{4}g_{pq}(\Ysp)\triangle_{{\rm X}}h_0=0,
  \label{eq:einstein-nw:b}
}
%
where $g_{\mu\nu}(\Xsp)$, $R_{\mu\nu}(\Xsp)$, 
$\triangle_{\rm X}$ and $D_{\mu}$ denote 
the metric tensor, the Ricci tensor, the Laplacian, 
and the covariant
derivative with respect to the metric $ds^2(\Xsp)$, 
respectively, and $g_{pq}(\Ysp)$ and $R_{pq}(\Ysp)$ 
denote the metric tensor and the Ricci tensor with 
respect to the metric $ds^2(\Ysp)$, respectively. 
Because $\triangle_{\rm X} h_0$ depends only on $x^\mu$, and 
$R_{pq}(\Ysp)$ and $g_{pq}(\Ysp)$ depend only on the 
coordinates $y^p$ of $\Ysp$, \eqref{eq:einstein-nw:b} 
requires that $\triangle_{\rm X} h_0$ is a constant. 
Hence, the equations \eqref{eq:einstein-nw} can be reduced to 
%
%
\Eqrsubl{eq:nf-condition}{
& &R_{\mu\nu}(\Xsp)=h_0^{-1}[D_{\mu}D_{\nu}h_0-\lambda g_{\mu\nu}(\Xsp)],
   \label{eq:nf-condition1}
\\
& &R_{pq}(\Ysp)=\lambda g_{pq}(\Ysp),
   \label{eq:nf-condition2}\\
&& \triangle_{\rm X}h_0=4\lambda,
   \label{eq:nf-condition3}
}
%
where $\lambda$ is a constant.

If we further assume that $\Xsp$ is Ricci flat, 
from \eqref{eq:nf-condition1}, the modulus $h_0$ is 
required to obey the equation
%
\begin{eqnarray}
\qquad D_{\mu}D_{\nu}h_0=\lambda g_{\mu\nu}(\Xsp).
  \label{eq:nwmoduli} 
\end{eqnarray}
%
In the case of $(D h_0)^2\not\equiv 0$, this equation 
has a solution only when $\Xsp$ is locally flat, 
and its general solution for $h_0$ is given by
%
\begin{eqnarray}
h_0(x)=\frac{\lambda}{2}x^{\mu} x_{\mu} + a_{\mu}x^{\mu}+b,
 \label{eq:nw-solution}
\end{eqnarray}
%
in terms of the four-dimensional Minkowski coordinates $x^{\mu}$, where 
$a_{\mu}$ and $b$ are constants satisfying the condition 
$a\cdot a\ne 0$. On the other hand, if $D_{\mu}h_0\ne 0$ 
and $(D_{\mu}h_0)^2=0$, there exists a solution only 
when $\lambda=0$, and a plane-wave-type geometry is 
also allowed for $\Xsp$ \cite{Kodama:2005fz}.

%
\subsection{Four-dimensional effective theory}
    \label{subsec:noflux-effective}

Next, we derive the four-dimensional effective theory 
for the four-dimensional spacetime and the size modulus 
in the setup of the previous subsection 
\ref{subsec:noflux-ansatz}, i.e., under the assumptions that 
the ten-dimensional metric is given by 
\eqref{eq:nw-general-metric}, the dilaton is constant, and  
all form fluxes vanish. We also require that the internal 
space $\Ysp$ has a fixed geometry satisfying \eqref{eq:nf-condition2}.

In this setup, the bosonic low-energy action for 
the ten-dimensional supergravity in the Einstein 
frame is simply given by the ten-dimensional Einstein-Hilbert action
\Eq{
S_{\rm IIB} = \frac{1}{2\tilde{\kappa}^2} \int_{\tXsp} d\Omega(\tXsp) 
    R(\tXsp), 
   \label{eq:10d-action}
}
where $\tilde{\kappa}$ is a positive constant. Here, 
under the assumption (\ref{eq:nw-general-metric}), the ten-dimensional 
scalar curvature $R(\tXsp)$ is expressed as
\begin{eqnarray}
R(\tXsp)=h_0^{1/2}R(\Xsp)+h_0^{-1/2}R(\Ysp)
   -\frac{3}{2}h_0^{-1/2}\triangle_{\rm X}h_0\,,
   \label{eq:nw-10d-Ricci}
\end{eqnarray}
where $R(\Xsp)$ and $R(\Ysp)$ are  the scalar curvatures of the 
metrics $ds^2(\Xsp)$ and $ds^2(\Ysp)$, respectively. 
Inserting this expression into the action (\ref{eq:10d-action}), 
we obtain the four-dimensional effective action
\Eq{
S_{\rm IIB}
 =\frac{1}{2\kappa^2} \int_{{\rm X}_4} d\Omega(\Xsp) 
  \left[h_0 R(\Xsp) + 6\lambda \right],
   \label{eq:nw4d-action}
}
where $\kappa$ is given by $\kappa=(V_6)^{-1/2}\tilde{\kappa}$, 
$V_6$ is the volume of the internal space ${\rm Y}_6$, 
\begin{equation}
V_6=\int_{{\rm Y}_6} d\Omega(\Ysp) \,,
   \label{eq:Calabi-volume}
\end{equation}
and we have dropped the surface term coming from 
$\triangle_{\rm X}h_0$. It is easy to see that the 
four-dimensional Einstein equations and the field 
equation for $h_0$ obtained from this effective action 
are exactly identical to \eqref{eq:nf-condition1} and 
\eqref{eq:nf-condition3}. Hence, the four-dimensional 
effective theory is equivalent to the original 
ten-dimensional theory under the ansatz adopted in this section.

Here, note that this effective theory has a kind of 
modular invariance when $\Ysp$ is a flat torus or a Calabi-Yau space. 
To see this, by the conformal transformation 
$ds^2(\Xsp)=h_0^{-1}ds^2(\bXsp)$, let us rewrite 
the four-dimensional effective action (\ref{eq:nw4d-action}) 
in terms of the metric $\bar g_{\mu\nu}$ in the Einstein frame as
\begin{eqnarray}
S_{\rm IIB}=\frac{1}{2\kappa^2} \int_{\bar{\rm X}_4} d\Omega(\bXsp) 
    \left[R(\bXsp) - \frac{3}{2}(\bar{D}\ln h_0)^2 
    +6\lambda h_0^{-2}\right],
   \label{eq:nwE4d-action}
\end{eqnarray}
where $R(\bXsp)$ and $\bar{D}_{\mu}$ are the  
Ricci scalar and the covariant derivative with 
respect to the metric $ds^2(\bXsp)$. It is easy to 
see that the action is invariant under the 
transformation $h_0\rightarrow h_0^{-1}$, provided 
that $\lambda=0$. Hence, if there is a solution 
for which $h_0$ increases in time, there is also 
a solution with the same four-dimensional metric 
in the Einstein frame such that $h_0$ decreases in time.

%
\section{Flux compactification in the 10D IIB supergravity}
    \label{sec:flux compactification}

In this section, we derive an effective theory describing 
the dynamics of the four-dimensional spacetime and the 
size modulus of the internal space for the flux 
compactification of the ten-dimensional type IIB 
supergravity.  Then, we study the difference in 
the spacetime dynamics and the behavior of the size 
modulus for  the four-dimensional effective theory 
and for the ten-dimensional theory.

%
\subsection{Ten-dimensional solutions}
   \label{subsec:IIB:10Dsol}

In our previous work \cite{Kodama:2005fz}, we derived 
a general dynamical solution for warped compactification 
with fluxes in the ten-dimensional type IIB supergravity. 
In that work, we imposed $d\dual(B_2\w H_3)=0$, which 
led to a slightly strong constraint on the free data for 
the solution, especially in the case of a compact 
internal space. Afterward, we have noticed that this 
condition is not necessary to solve the field equations, 
and without that condition, we can find a more general 
class of solutions. Because we take this class as the 
starting point of our argument, we first briefly 
explain how to get a general solution without that condition. 
We omit the details of calculations because they are 
essentially contained in our previous paper \cite{Kodama:2005fz}.

We assume that the ten-dimensional spacetime metric takes the form
\Eq{
ds^2(\tXsp)=A(x,y)^2 ds^2(\Xsp)+B(x,y)^2ds^2(\Ysp),
\label{IIB:metric:general}
}
where the meanings of $ds^2(\Xsp)$ and $ds^2(\Ysp)$ 
and the other related notations are the same as in 
the previous section. $A(x,y)$ and $B(x,y)$ are 
arbitrary non-vanishing functions on $\tXsp$ at 
the beginning. We further require that the dilaton 
and the form fields satisfy the following conditions:
\Eqrsubl{IIB:gauge-assumptions}{
& &\tau\equiv C_0+i\, e^{-\Phi}=ig_s^{-1}(=\const)\,,
   \label{IIB:gauge-assumption1}
      \\
& &G_3\equiv ig_s^{-1}\,H_3-F_3
  =\frac{1}{3!}\,G_{pqr}(y)\,dy^p\wedge dy^q \wedge dy^r\,,
   \label{IIB:gauge-assumption2}
      \\
& &\ast_{\rm Y}\,G_3= \epsilon i\,G_3\quad(\epsilon=\pm1)\,,
  \label{IIB:gauge-assumption2-1}
      \\
& & \tilde F_5
   =(1\pm\dual){V_p dy^p} \wedge \Omega(\Xsp)
   ={V}\wedge \Omega(\Xsp) \mp A^{-4}B^4 *_{\rm Y} V,
  \label{IIB:gauge-assumption4}
}
where $g_s$ is a constant representing the string 
coupling constant, and $\ast$ and $\ast_{\rm Y}$ are 
the Hodge duals with respect to the ten-dimensional 
metric $ds^2(\tXsp)$ and the six-dimensional metric 
$ds^2(\Ysp)$, respectively. 

Under these assumptions, the two of the field equations,
\Eqrsub{
&& \Box \tau + i\frac{(\nabla\tau)^2}{\tau_2}
  =-\frac{i}{2}G_3\cdot G_3,\\
\label{IIB:FEq:tau}
&& \dual \tilde F_5=\pm \tilde F_5,
\label{IIB:FEq:F5:duality}
}
are automatically satisfied, and the rest are written
\Eqrsub{
&& dG_3=0,
\label{IIB:FEq:G3:1}\\
&& \nabla\cdot G_3=* d * G_3=-i G_3\cdot\tilde F_5,
\label{IIB:FEq:G3:2}\\
&& d \tilde F_5=H_3\wedge F_3,
\label{IIB:FEq:dF5}\\
&& R_{MN}=\frac{g_s}{4} \insbra{\Re(G_{MPQ}G^*_N{}^{PQ})
  -\frac{1}{2}G_3\cdot G_3^* g_{MN} }
  +\frac{1}{96} \tilde F_{MP_1\cdots P_4}\tilde F_N{}^{P_1\cdots P_4}.
\label{IIB:FEq:Einstein}
}
%

Among these equations, the first together with the 
assumptions \eqref{IIB:gauge-assumption2} and 
\eqref{IIB:gauge-assumption2-1} implies that $G_3$ is 
a closed imaginary-self-dual (ISD) 3-form on $\Ysp$ 
that does not depend on the coordinates $x^\mu$. 
Then, \eqref{IIB:FEq:G3:2} can be rewritten as
\Eq{
\inpare{V \mp \epsilon d_y(A^4)}\cdot G_3=0,
}
where $d_y =dy^p \pd_p$. Since $G_3$ is an ISD form on 
$\Ysp$, and $V$ and $d_y(A^4)$ are 1-forms on $\Ysp$, 
it follows from this equation that
\Eq{
V=\pm \epsilon d_y (A^4),
}
provided $G_3\not=0$.  

Inserting this expression into \eqref{IIB:FEq:dF5}, 
we obtain the following two equations:
\Eqrsub{
&& \pd_\mu (A^{-4}B^4 \pd_p (A^4))=0,
\label{IIB:FEq:dF5:1}\\
&& (\hat D\cdot( A^{-4}B^4 \hat D (A^4))_{\rm Y}
    =\frac{g_s}{2} (G_3\cdot \bar G_3)_{\rm Y},
\label{IIB:FEq:dF5:2}
}
where $\hat D_p$ is the covariant derivative with 
respect to the metric $ds^2(\Ysp)$, and 
$(\alpha\cdot\beta)_{\rm Y}$ denotes the inner product of forms 
$\alpha$ and $\beta$ on $\Ysp$ with respect to the metric 
$ds^2(\Ysp)$.

Next, we consider the Einstein equations. 
First, from $R_{ap}=0$ and \eqref{IIB:FEq:dF5:1}, we find that we can set 
\Eq{
A=h(x,y)^{-1/4},\quad B=h(x,y)^{1/4},
\label{ABbyh}
}
by appropriately redefining $ds^2(\Xsp)$ and 
$ds^2(\Ysp)$. Correspondingly, $\tilde F_5$ and 
\eqref{IIB:FEq:dF5:2} can be written as
\Eqr{
&& \tilde F_5= \pm\epsilon(1\pm\dual) d(h^{-1}) \w \Omega(\Xsp),
\label{F5byh}\\
&& \triangle_{\rm Y} h = -\frac{g_s}{2}(G_3\cdot \bar G_3)_{\rm Y}.
\label{hbyG3}
}

With these expressions, the ten-dimensional 
Einstein equations \eqref{IIB:FEq:Einstein} read
\Eqrsubl{IIB:W-Einstein}{
& & h R_{\mu\nu}(\Xsp)- D_{\mu}D_{\nu}h
 +\frac{1}{4}g_{\mu\nu}(\Xsp)\triangle_{\rm X}h=0, \\
     \label{IIB:W-Einstein1} 
& &\pd_{\mu}\pd_{p}h=0,\\
     \label{IIB:W-Einstein2}  
& & R_{pq}(\Ysp)-\frac{1}{4}g_{pq}(\Ysp) \triangle_{\rm X}h=0.
     \label{IIB:W-Einstein3} 
}
{}From the second of these equations, we immediately 
see that the warp factor $h$ can be expressed as
\Eq{
h(x,y)=h_0(x)+h_1(y).
\label{IIB:sol:general:h}
}
Further, if we require that $d_y h \ne 0$, 
the rest of the equations can be reduced to
\Eqrsubl{IIB:sol}{
&& R_{\mu\nu}(\Xsp)=0, 
 \label{IIB:sol:X}\\
&& D_{\mu}D_{\nu}h_0= \lambda g_{\mu\nu}(\Xsp),
 \label{IIB:sol:h0}\\
&& R_{pq}(\Ysp)=\lambda g_{pq}(\Ysp). 
 \label{IIB:sol:Y}
}
%

Thus, we have found that the most general solutions 
satisfying the conditions \eqref{IIB:metric:general} 
and \eqref{IIB:gauge-assumptions} are specified by a 
Ricci flat spacetime $\Xsp$, an Einstein space $\Ysp$, 
a closed ISD 3-form $G_3$ on $\Ysp$, and the function 
$h(x,y)$ that is the sum of $h_0(x)$ satisfying 
\eqref{IIB:sol:h0} and $h_1(x)$ satisfying \eqref{hbyG3}. 
The additional constraint on $G_3$, 
$d_y[h^{-2}(B_2\cdot dB_2)_{\rm Y}]=0$, in Ref. \cite{Kodama:2005fz} 
does not appear. Further, closed ISD 3-forms on $\Ysp$ 
are in one-to-one correspondence with real harmonic 3-forms on 
$\Ysp$.  Hence, this class of dynamical solutions exist 
even for a generic compact Calabi-Yau internal space, 
if we allow $h_1(y)$ to be a singular function. This 
singular feature of $h$ in the compact case with flux 
arises because $h$ is a solution to the Poisson equation 
\eqref{hbyG3} and has nothing to do with the dynamical 
nature of the solution. It is shared by the other flux 
compactification models.

Here, note that the Ricci flatness of $\Xsp$ is required 
from the Einstein equations. This should be contrasted 
with the previous case with no warp. This point is quite 
important in the effective theory issue, as we see soon. 
Anyway, as explained in the previous section, the Ricci 
flatness of $\Xsp$ and \eqref{IIB:sol:h0} are consistent 
only when $\Xsp$ is locally flat if $(Dh_0)^2\not=0$.

%
\subsection{Four-dimensional effective theory} 
 \label{subsec:flux-effective}
 
Now we study the four-dimensional effective theory that 
incorporates the dynamical solutions obtained in the 
previous subsection. For simplicity, we do not consider 
the internal moduli degrees of freedom of the metric of 
$\Ysp$ or of the solution $h_1(y)$ in the present paper. 
Then, in its $x$-independent subclass with $\lambda=0$, 
we have only one free parameter $h_0$. When we rescale  
$ds^2(\Ysp)$ by a constant $\ell$ as $\ell^2 ds^2(\Ysp) 
\maps ds^2(\Ysp)$, we have to rescale $h$ as $h/\ell^4 
\maps h$. We can easily see that the corresponding rescaled 
$h_1$ satisfies  \eqref{hbyG3} again with the same $G_3$ as 
that before the rescaling. We can also confirm that the D3 
brane charges associated with the 5-form flux do not change 
by this scaling. In contrast, $h_0$ changes its value by 
this rescaling. Therefore, $h_0$ represents the size 
modulus of the Calabi-Yau space $\Ysp$. 

{}From this observation, we construct the four-dimensional 
effective theory for the class of ten-dimensional 
configurations specified as follows. First, we assume 
that $\tXsp$ has the metric 
\Eq{
ds^2(\tXsp) = h^{-1/2}(x,y)\,ds^2(\Xsp) +h^{1/2}(x,y)\,ds^2(\Ysp),
   \label{IIB:metric}
}
where $h=h_0(x)+h_1(y)$ and $ds^2(\Ysp)$ is a fixed 
Einstein metric on $\Ysp$ satisfying \eqref{IIB:sol:Y}, 
while $ds^2(\Xsp)$ is an arbitrary metric on $\Xsp$. 
Further, we assume that the dilaton is frozen as in 
\eqref{IIB:gauge-assumption1}, $G_3$ is given by a fixed 
closed ISD 3-form on $\Ysp$, $h_1(y)$ is a fixed solution 
to \eqref{hbyG3}, and $\tilde F_5$ is given by \eqref{F5byh}. 
Hence, the metric of $\Xsp$ and the function $h_0$ on it 
are the only dynamical variables in the effective theory.

The four-dimensional effective action for these variables 
can be obtained by evaluating the ten-dimensional action 
of the IIB theory
\Eqr{
S_\r{IIB} &=& \frac{1}{2\tilde\kappa^2}\int_{\tXsp}d\Omega(\tXsp)
 \insbra{ R(\tXsp)-\frac{\nabla_M \bar\tau \nabla^M \tau}{2(\Im\tau)^2}
   -\frac{G_3\cdot\bar G_3}{2\Im\tau}-\frac{1}{4}\tilde F_5^2}
   \notag\\
   &&\pm \frac{i}{8\tilde \kappa^2}\int_{\tXsp} 
   \frac{C_4\wedge G_3\wedge \bar G_3}{\Im\tau},
\label{IIB:action}}
for the class of configurations specified above. In general, 
there is subtlety concerning the action of the type IIB 
supergravity, because the correct field equations can be 
obtained by imposing the self-duality condition 
\eqref{IIB:FEq:F5:duality} after taking variation of the 
action in general. In the present case, however, since 
we are only considering configurations \eqref{F5byh} 
satisfying the self-duality condition, this problem 
does not affect our argument. We can obtain the "correct" 
effective action by simply inserting \eqref{F5byh} into 
the above ten-dimensional action. 

First, for the metric \eqref{IIB:metric} with 
$h=h_0(x)+h_1(y)$, the ten-dimensional scalar curvature 
$R(\tXsp)$ is expressed as
\Eqr{
R(\tXsp)=h^{1/2}R(\Xsp)+h^{-1/2}R(\Ysp)
   -\frac{3}{2}h^{-1/2}\triangle_{\rm X}h_0
   -\frac{1}{2}h^{-3/2}\hat{\triangle}_{\rm Y}h_1,
   \label{eq:10d-Ricci}
}
where $\triangle_{\rm X}$ and $\hat{\triangle}_{\rm Y}$ 
are the Laplacian with respect to the metrics $ds^2(\Xsp)$ 
and $ds^2(\Ysp)$, respectively. Inserting this expression, 
\eqref{F5byh}, \eqref{hbyG3} and \eqref{IIB:sol:Y} into 
\eqref{IIB:action}, we get 
\Eqr{
S_{\rm IIB}&=&\frac{1}{2\kappa^2} \int_{{\rm X}_4} d\Omega(\Xsp)
\left[H(x) R(\Xsp)+6\lambda+\frac{1}{2V_6}\int_{{\rm Y}_6} d\Omega(\Ysp)
    \,h^{-1}\hat{\triangle}_{{\rm Y}}h_1 \right]\nn\\     
   & & \pm \frac{i}{8\tilde{\kappa}^2}\int_{\tXsp} 
     \frac{C_{(4)}\wedge G_{(3)}
     \wedge \bar{G}_{(3)}}{{\rm Im}\tau},
   \label{eq:4d-effectiveaction}
}
where we have dropped the surface term coming from 
$\triangle_{\rm X} h_0$, $\kappa=(V_6)^{-1/2}\tilde{\kappa}$, 
and $H(x)$ is defined by 
\Eq{
H(x)=h_0(x)+ c;\quad c:=V_6^{-1}\int_{{\rm Y}_6}d\Omega(\Ysp)h_1.
}
The Chern-Simons term in this expression can be rewritten 
as follows. First, \eqref{hbyG3} can be written 
\Eq{
i\epsilon g_s G_3\w \bar G_3=2 d(\dual_{\rm Y} dh_1).
}
{}From this, it follows that
\Eq{
ig_s C_4\w G_3 \w \bar G_3=d(2\epsilon C_4\w \dual_{\rm Y} dh_1)
    \mp 2h^{-2} (dh_1\cdot dh_1)_{\rm Y}\,\Omega(\Xsp)\wedge\Omega(\Ysp).
}
Hence, we have
\Eqr{
&& \pm \frac{i}{8\tilde{\kappa}^2}\int_{\tXsp} \frac{C_{4}\wedge G_{3}
     \wedge \bar{G}_{3}}{{\rm Im}\tau}
     =
     -\frac{1}{4V_6\kappa^2}
     \int_{{\rm X}_4} d\Omega(\Xsp)
     \int_{{\rm Y}_6} d\Omega(\Ysp)\,
     \frac{\hat\triangle_{\rm Y} h_1}{h} 
     \notag\\
&&\qquad
   +\frac{1}{4\kappa^2 V_6}\int_{\tXsp} 
       d\insbra{\inpare{\pm\epsilon C_4
        -h^{-1} \Omega(\Xsp)} \w \dual_{\rm Y} dh_1} \,.
}
Note that apart from the boundary term,  the contribution 
of Chern-Simons term is canceled by the term containing $h_1$ in  \eqref{eq:4d-effectiveaction}, which came from 
the ten-dimensional scalar curvature and the 3-form flux. 
Consequently, neglecting the boundary term, we obtain the 
following four-dimensional effective action 
\Eq{
S_{\rm IIB}=\frac{1}{2\kappa^2} \int_{{\rm X}_4} d\Omega(\Xsp) 
   \left[H R(\Xsp) +6\lambda \right].
   \label{eq:4d-action}
}
%

This effective action has the same form as 
\eqref{eq:nw4d-action}. Hence, it gives the 
four-dimensional field equations of the same form
 as in the no-flux case:
\Eqrsubl{eq:4d-Eequation}{
& &R_{\mu\nu}(\Xsp)=H^{-1} 
   \left[D_{\mu} D_{\nu} H - \lambda g_{\mu\nu}(\Xsp)\right],
\label{eq:4d-Eequation1}
\\
& &\triangle_{\rm X} H= 4\lambda.
\label{eq:4d-equation2}
}
If the four-dimensional spacetime is Ricci flat, these 
equations reproduce the correct equation for $h_0(x)=H-c$ 
obtained from the ten-dimensional theory in the previous 
subsection. However,  the Ricci flatness of $\Xsp$ is not 
required in the effective theory unlike in the 
ten-dimensional theory. Hence, the class of solutions 
allowed in the four-dimensional effective theory is much 
larger than the original ten-dimensional theory. 

In particular, the effective theory has a modular invariance 
similar to that found in the no-flux Calabi-Yau 
case with $\lambda=0$. In fact, by the conformal 
transformation $ds^2(\Xsp)=H^{-1}ds^2(\bXsp)$, 
\eqref{eq:4d-action} is expressed in terms of the 
variables in the Einstein frame as 
\begin{eqnarray}
S_{\rm IIB}=\frac{1}{2\kappa^2} \int_{\bar{\rm X}_4} d\Omega(\bXsp)
 \left[R(\bXsp) - \frac{3}{2} 
 (\bar{D}\ln H)^2 +6\lambda H^{-2}\right],
   \label{eq:E4d-action}
\end{eqnarray}
where $R(\bXsp)$ and $\bar{D}_{\mu}$ are  the scalar 
curvature and the covariant derivative with respect to 
the metric $ds^2(\bXsp)$. 
The corresponding four-dimensional Einstein equations 
in the Einstein frame and the field equation for $H$ 
are given by 
\Eqrsubl{eq:E4d-equation}{
& &R_{\mu\nu}(\bXsp)=\frac{3}{2}\bar{D}_{\mu}\ln H\, 
   \bar{D}_{\nu}\ln H - 3\lambda H^{-2}g_{\mu\nu}(\bXsp),\\
& &\triangle_{\bar{\rm X}}\ln H= 4\lambda H^{-2},
}
where $\triangle_{\bar{\rm X}}$ is the Laplacian with 
respect to the metric $ds^2(\bXsp)$. 
It is clear that for $\lambda=0$, this action and the 
equations of motion are invariant under the transformation 
$H\rightarrow k/H$, where $k$ is an arbitrary positive 
constant. 

This transformation corresponds to the following 
transformation in the original ten-dimensional metric. 
Let us denote the new metric of $\Xsp$ and the function 
$h$ obtained by this transformation by $ds'{}^2(\Xsp)$ 
and $h'$, respectively. Then, since the transformation 
preserves the four-dimensional metric in the Einstein 
frame, $ds'{}^2(\Xsp)$ is related to $ds^2(\Xsp)$ as 
$ds'{}^2(\Xsp)=(H^2/k) ds^2(\Xsp)$. In the meanwhile, 
from $H'=k/H= h_0' + c$, $h'$ is expressed in terms of 
the original $h_0$ as
\Eq{
h'= \frac{k}{h_0(x)+c}- c + h_1(y).
}
The corresponding ten-dimensional metric is written
\Eq{
ds^2= k^{-1}H^2 (h')^{-1/2} ds^2(\Xsp)+ (h')^{1/2}ds^2(\Ysp).
}
It is clear that this metric and $h'$ do not satisfy the 
original ten-dimensional field equations. Hence, the 
modular-type invariance of the four-dimensional effective 
theory is not the invariance of the original ten-dimensional 
theory.


\section{Ho\v{r}ava-Witten model in the 11D heterotic M-theory}
 \label{sec:Horava-Witten}

A dynamical solution similar to that of the ten-dimensional 
IIB discussed in the previous section was found by 
Chen et al. \cite{Chen:2005jp} for the five-dimensional 
effective theory obtained from the Ho\v{r}ava-Witten 
model of the eleven-dimensional M-theory. In this section, 
we derive a four-dimensional effective theory for this 
five-dimensional theory.

\subsection{Five-dimensional effective theory}

We first briefly summarise the argument leading to the 
five-dimensional effective theory for the Ho\v{r}ava-Witten 
model of the eleven-dimensional M-theory. In this model, 
we first compactify the M-theory in eleven dimensions over 
$S^1/\ZR_2$. Let the length of this compactifying circle 
$S^1$ be $2L$. Then, it is expected that $E_8$ gauge fields 
and their superpartners are created on the two orientifold 
planes to cancel the anomalies, leading to the 
$E_8\times E_8$ heterotic theory in ten dimensions 
in the limit of small $L$. Hence, the action of the 
Ho\v rava-Witten model is given by 
\cite{Horava:1995qa,Horava:1996ma}
\Eqr{
S_{\rm HW}&=&\frac{1}{2\hat{\kappa}^2}
   \int_{\hat{{\rm X}}_{11}} d\Omega(\hat{{\rm X}}_{11})
   \left[R(\hat{{\rm X}}_{11})-\frac{1}{2}F_4^2\right]
   -\frac{1}{12\hat{\kappa}^2}\,A_3\,\wedge\,F_4\,\wedge\,F_4\nn\\
 & &
  -\frac{1}{8\pi\hat{\kappa}^2}
   \left(\frac{\hat{\kappa}}{4\pi}\right)^{2/3}
  \sum_{j=1,2}\int_{{\rm X}^{(j)}_{10}} d\Omega({\rm X}_{10})
\left[{\rm tr}(F^{(j)})^2-\frac{1}{2}{\rm tr} R^2\right],
  \label{eq:HWaction}
}
where $\hat{\kappa}$ is the positive constant, 
$R(\hat{{\rm X}}_{11})$ is the scalar curvature with respect to the 
eleven-dimensional metric $ds^2(\hat{{\rm X}}_{11})$, 
$A_3$ is the 3-form gauge field with the field strength 
$F_4=dA_3$, and $F^{(1)}$ and $F^{(2)}$ are the $E_8$ 
gauge field strengths. We choose the range $-L \le z \le L$ 
for the coordinate of $S^1$ with the end points being 
identified and impose the $\ZR_2$ symmetry under the 
transformation $z \maps -z$. The orientifold planes 
of this transformation, ${\rm X}_{10}^{(i)} (i=1, 2)$, 
correspond to $z=0$ and $z=L$. For simplicity, we will 
not consider the boundary gauge fields in the present paper.

If we further compactify this model over a six-dimensional 
internal space $\Ysp$, then we obtain a four-dimensional 
model. In practice, the argument becomes simpler if we 
reverse the order of compactifications, i.e., if we 
compactify the M-theory first over $\Ysp$ to 
$\hat{\rm X}_{11}=\tilde {\rm X}_5 \times \Ysp$ and then over 
$S^1/\ZR_2$ to $\tilde {\rm X}_5=\Xsp\times S^1/\ZR_2$, as was 
done by Lukas et al \cite{Lukas:1998yy}. In the first step, 
we obtain an effective five-dimensional theory. At this step, 
we assume that the eleven-dimensional metric takes the form
\Eq{
ds^2(\hat{{\rm X}}_{11})=e^{2\phi(\tilde{x})/3}ds^2(\tilde{\rm X}_5)
   +e^{-\phi(\tilde{x})/3}ds^2(\Ysp),
  \label{eq:HWmetric}
}
where $\tilde{x}^a$ are the coordinates of the 
five-dimensional spacetime $\tilde{\rm X}_5$, and 
that the 4-form flux is expressed as 
\Eq{
F_4=(\omega\cdot\Omega(\Ysp))_{\rm Y}\,,
    \label{eq:HWflux}
}
where $\omega$ is a 2-form on $\Ysp$. We can show that 
even if we start from a more general warped metric of 
the form 
$ds^2(\hat{{\rm X}}_{11})=e^{\alpha}ds^2(\tilde {\rm X}_5)
+e^\beta ds^2(\Ysp)$, 
the field equations require both $\alpha$ and $\beta$ 
to depend either only on $\tilde x^a$ or only on $y^p$, 
if $F_4$ takes the form \eqref{eq:HWflux}. Hence, the 
above choice for the metric form is quite natural when 
we study dynamical instability of supersymmetric solutions 
in the Ho\v{r}ava-Witten model. 

{}From the field equations 
\Eq{
d F_4=0,\quad
d\dual F_4+ \frac{1}{2} F_4 \w F_4=0,
}
we obtain
\Eq{
\pd_a \omega=0,\quad
d\omega=0,\quad 
\hat D\cdot \hat\omega=0,
}
where $\hat{\omega}$ is a 2-form on $\Ysp$ such that 
$\hat\omega_{pq}=\omega_{pq}$ and their indices are 
raised and lowered by the metric $ds^2(\Ysp)$. Next, 
the traceless part of the Einstein equations for $R_{pq}$ gives

\Eq{
 R_{pq}(\Ysp)-\frac{1}{6} R(\Ysp) \hat g_{pq}(\Ysp)
  =-\frac{e^{\phi}}{2}
   \inpare{\hat\omega_{pr}\hat\omega_q{}^r
    -\frac{1}{3}\hat\omega^2 \hat g_{pq}}.
}
{}From this, it follows that if $\pd_a\phi\not=0$, 
both sides of this equation should vanish separately. 
Hence, taking account of the Bianchi identity, we obtain
\Eqr{
&& R_{pq}(\Ysp)=\lambda g_{pq}(\Ysp),\\
&& \hat\omega_{pr}\hat\omega_q{}^r
         =\frac{1}{3}\hat\omega^2 g_{pq}(\Ysp).
}
Inserting these relations to the $R^p_p$ equation, we have
\Eq{
6 \lambda +e^{-\phi} \tilde D^2\phi
  = e^{\phi}\hat\omega^2,
}
from which and the rest of the Einstein equations, we obtain 
the constraint $\hat\omega^2=2m^2=\const$ and the field 
equations in the five-dimensional theory
\Eqrsubl{HW5D:FEqs}{
&& {R}_{ab}(\tilde{\rm X}_5)
       =\frac{1}{2}\pd_a \phi \pd_b \phi
      + \inpare{\frac{m^2}{3}e^{2\phi}-2 \lambda e^\phi}
              g_{ab}(\tilde{\rm X}_5),
   \label{HW5D:FEq:Einstein}\\
&& \Box_{\tilde{\rm X}}\phi-2m^2e^{2\phi}= -6\lambda e^\phi,
  \label{HW5D:FEq:dilaton}
}
where $\Box_{\tilde{\rm X}}$ is the D'Alermbertian for 
the five-dimensional metric $ds^2(\tilde{\rm X}_5)$.

These field equations can be obtained from  the 
five-dimensional effective action given by 
\cite{Chen:2005jp,Lukas:1998yy}
\Eq{
S_{\rm HW}=\frac{1}{2\tilde{\kappa}^2}
         \int_{\tilde{\rm X}_5} d\Omega(\tilde{\rm X}_5) 
 \left[R(\tilde{\rm X}_5)-\frac{1}{2}(\tilde{D}\phi)^2
          -m^2 e^{2\phi}+ 6 \lambda e^\phi\right],
    \label{HW5D:5Daction}
}
where $\tilde{\kappa}=(V_6)^{-1/2}\hat{\kappa}$, 
$V_6$ is the volume of $\Ysp$.

\subsection{Four-dimensional effective theory}

In the Ho\v{r}ava-Witten model, a four-dimensional theory 
is obtained from the five-dimensional theory by 
compactification over $S^1/\ZR_2$. Without loss of generality, 
the metric obtained by this compactification can be 
written $ds^2=e^\gamma ds^2(\Xsp)+ e^\delta dz^2$. 
In general, the field equations do not lead to no 
relation between the warp factors $e^\gamma$ and 
$e^\delta$ in this theory, and there exists no natural 
reduction to four dimensions. Hence, in order to obtain 
a four dimensional reduction, we have to impose some 
relation between $e^\gamma$ and $e^\delta$. In the present 
paper, to  include the dynamical solution found by 
Chen et al. \cite{Chen:2005jp}, we adopt the ansatz that 
$ds^2(\tilde {\rm X}_5)$ can be written
\Eq{
 ds^2(\tilde{\rm X}_5)=h^{1/2}(x, z)\,ds^2(\Xsp)+h(x, z)\,dz^2, 
   \label{eq:ht-solution}
}
and the warp factor $h$ has the structure 
\Eq{
h(x, z)=h_0(x)+ h_1(z).
 \label{HW5D:Ansatz}
} 
We also assume that $\Ysp$ is a Calabi-Yau space, i.e. $\lambda=0$. 
As is shown in Appendix \ref{Appendix:horava}, the most 
general solution to the field equations \eqref{HW5D:FEqs} 
satisfying this ansatz and the conditions 
$\pd_\mu h_0\not=0$ and $\pd_z h_1\not=0$ is given by 
\Eq{
R_{\mu\nu}(\Xsp)=0,\qquad
h(x, z)=h_0(x)+ k z, \qquad e^{2\phi}=h^{-3},
    \label{HW5D:HWsolution}
}
where $k^2=8m^2/3$, and $h_0$ is a solution to 
\Eq{
D_\mu D_\nu h_0=0.
\label{HW5D:HWsolution:addCond}
}
In the case $(Dh_0)^2\not=0$, which requires that $\Xsp$ 
is locally flat \cite{Kodama:2005fz}, this solution 
(\ref{HW5D:HWsolution}) is identical to the solution 
found by Chen et al. \cite{Chen:2005jp} 
(See Appendix \ref{Appendix:horava}).

On the basis of this result, we construct a four-dimensional 
effective theory of the Ho\v{r}ava-Witten model for the 
class of five-dimensional configurations in which the 
metric is expressed as \eqref{eq:ht-solution} with $h$ 
of the form \eqref{HW5D:Ansatz}, and $\phi$ is related to $h$ by
\Eq{
\phi=-\frac{3}{2}\ln h.
}
For this class of configurations, the five-dimensional 
action (\ref{HW5D:5Daction}) can be written 
\Eq{
S_{\rm HW}=\frac{1}{2\tilde{\kappa}^2}\int_{\Xsp} 
    d\Omega(\Xsp) \int^L_{0} dz 
    \left[h\,R(\Xsp) -\frac{2\pd_z^2 h_1}{h^{1/2}} 
    +\frac{5(\pd_z h_1)^2}{8 h^{3/2}} - \frac{3k^2}{8 h^{3/2}}
    \right].
   \label{eq:gb-faction}
}
In order to perform the integration over $z$, we have 
to specify $h_1(z)$. In the present case, the only possible choice is 
\Eq{
h_1(z)=k z;\quad k^2=\frac{8m^2l^2}{3}.
}
However, the simple insertion of this expression into 
the above action does not give a correct result. This 
is because the variation of the action 
\eqref{eq:gb-faction} with respect to $h$ produces boundary 
terms at the orientifold planes $z=0,\,L$, which do not 
vanish for the above choice of $h_1$. By inspecting the 
structure of these boundary terms, we find that if we add 
the additional term to the action given by
\Eq{
S_{\rm boundary}=\frac{1}{2\tilde{\kappa}^2}\int_{{\rm X}_4}
d\Omega(\Xsp)
  \left[\frac{1}{2}h^{-1/2}k\right]^{z=L}_{z=0},
   \label{eq:boundary}
}
the correct field equations are obtained in five-dimension. 
Therefore, the four-dimensional effective action is given by
\Eq{
S \equiv S_{\rm HW}+S_{\rm boundary}\nn\\
    =\frac{1}{2\kappa^2}\int_{{\rm X}_4} d\Omega(\Xsp) H(x) R(\Xsp),
   \label{eq:gb-4daction}
}
where $\kappa=(L)^{-1/2}\tilde{\kappa}$, and $H(x)$ is defined by
\Eq{
H(x)=h_0(x)+\frac{kL}{2}\,.
}

Thus, we have obtained the same four-dimensional 
effective action as in the case of the type IIB 
supergravity in ten dimensions. In particular, the 
four-dimensional effective theory of the Ho\v{r}ava-Witten 
model allows solutions that cannot be uplifted to 
solutions in five dimensions or in eleven dimensions and 
has the same modular invariance as in the previous case, 
which is not respected in the original higher-dimensional 
theory, with respect to the size modulus in the Einstein frame.

\section{Conclusion}
\label{sec:conclusion}

In the present paper, we have derived four-dimensional 
effective theories for the spacetime metric and the size 
modulus of the internal space for warped compactification 
with flux in the ten-dimensional type IIB supergravity 
and in the Ho\v{r}ava-Witten model of the eleven-dimensional 
M-theory. The basic idea was to consider field configurations 
in higher dimensions that are obtained by replacing the 
constant size modulus in supersymmetric solutions for 
warped compactifications, by a field on the four-dimensional 
spacetime. The effective action for this moduli field and 
the four-dimensional metric has been determined by 
evaluating the higher-dimensional action for such 
configurations. In all cases, the dynamical solutions 
in the ten- and eleven-dimensional theories found by 
Gibbons et al. \cite{Gibbons:2005rt}, Kodama and 
Uzawa \cite{Kodama:2005fz} and Chen et al. \cite{Chen:2005jp} 
were reproduced in the four-dimensional effective theories. 

In addition to this, we have found that these 
four-dimensional effective theories have some unexpected 
features. First, the effective actions of both theories 
are exactly identical to the four-dimensional effective 
action for direct-product type compactifications with no 
flux in ten-dimensional supergravities. In particular, 
the corresponding effective theory has a kind of modular 
invariance with respect to the size modulus field in the 
Einstein frame. This implies that if there is a solution 
in which the internal space expands with the cosmic expansion, 
there is always a conjugate solution in which the internal 
space shrinks with the cosmic expansion. 

Second, the four-dimensional effective theory for warped 
compactification allows solutions that cannot be obtained 
from solutions in the original higher-dimensional theories. 
The modular invariance in the four-dimensional theory 
mentioned above is not respected in the original 
higher-dimensional theory either. This situation should be 
contrasted with the no-warp case in which the four-dimensional 
effective theory and the original higher-dimensional theory 
are equivalent under the product-type ansatz for the metric 
structure. This result implies that we have to be careful 
when we use a four-dimensional effective theory to analyse 
the moduli stabilisation problem and the cosmological 
problems in the framework of warped compactification of 
supergravity or M-theory.

\section*{Acknowledgments}
      
The authors would like to thank Renata Kallosh, Andrei Linde, 
Jiro Soda and Kazuya Koyama for valuable discussions. 
K.~U. is grateful to Misao Sasaki for continuing 
encouragement. 
This work is supported by the JSPS grant No. 15540267 
(H.~K.) and by the Yukawa fellowship (K.~U.).

\section*{Appendix}
\appendix

%
\section{Solutions of the 5D Ho\v rava-Witten model}
\label{Appendix:horava}

In this appendix, we prove that the solutions specified 
by \eqref{HW5D:HWsolution} and \eqref{HW5D:HWsolution:addCond} 
exhaust all solutions to the field equations \eqref{HW5D:FEqs} 
in the five-dimensional Ho\v{r}ava-Witten theory, if we 
assume that the five-dimensional metric takes the form
\Eq{
ds^2(\tilde{\rm X}_5)=h^{1/2}\,ds^2(\Xsp)+h\,dz^2,
   \label{HW5D:ap-solution}
}
with
\Eq{
h = h_0(x) + h_1(z);\qquad 
\pd_\mu h_0\not=0,\quad \pd_z h_1\not=0.
 \label{HW5D:ap-Ansatz}
}
%

For the metric \eqref{HW5D:ap-solution}, the field equations 
\eqref{HW5D:FEqs} can be written
\Eqrsub{
&& \!\!\!\!\!\!\!  D\cdot(h D\phi)+ \pd_z (h^{1/2}\pd_z \phi)
   =2m^2 h^{3/2} e^{2\phi},
     \label{HW5D:DilatonEq}\\
&& \!\!\!\!\!\!\!  R_{\mu\nu}(\Xsp)-\frac{1}{4}g_{\mu\nu}(\Xsp)R(\Xsp)
  + \frac{9}{8h^2}\left[D_\mu h D_\nu h
     -\frac{1}{4}(Dh)^2 g_{\mu\nu}(\Xsp)\right]
  \notag\\
&& 
   -\frac{1}{h}\left[D_\mu D_\nu h 
   -\frac{1}{4} \triangle_{\rm X}h g_{\mu\nu}(\Xsp)\right]
   =\frac{1}{2}\left[D_\mu \phi D_\nu\phi
     -\frac{1}{4} (D\phi)^2 g_{\mu\nu}(\Xsp)\right],
     \label{HW5D:EinsteinEqs:RabOD}\\
&& \!\!\!\!\!\!\!
   R({\rm X}_4)-\frac{2\triangle_{\rm X}h}{h} + \frac{9(Dh)^2}{8h^2}
   -\frac{\pd_z^2h}{h^{3/2}}+\frac{(\pd_z h)^2}{2h^{5/2}}
  =\frac{1}{2}(D\phi)^2 + \frac{4}{3}m^2 e^{2\phi} h^{1/2},
     \label{HW5D:EinsteinEqs:RabTr}\\
&& \!\!\!\!\!\!\!  
   -\frac{3}{4} h^{1/2} D_\mu\left(\frac{\pd_z h}{h^{3/2}}\right)
   =\frac{1}{2}\pd_\mu \phi \pd_z \phi,
     \label{HW5D:EinsteinEqs:Ray}\\
&& \!\!\!\!\!\!\!
   -\frac{ \triangle_{\rm X} h}{2h^{1/2}} -\frac{\pd_z^2 h}{h}
    +\frac{5 (\pd_z h)^2}{4h^2}
    = \frac{1}{2} (\pd_z \phi)^2 + \frac{1}{3} m^2 e^{2\phi} h,
     \label{HW5D:EinsteinEqs:Ryy}
}
where $R(\Xsp)$, $R_{\mu\nu}(\Xsp)$, $\triangle_{\rm X}$ 
and $D_{\mu}$ are the scalar curvature, the Ricci tensor, 
the Laplacian and the covariant derivative with respect 
to the metric $ds^2(\Xsp)$. 

First, from the assumption \eqref{HW5D:ap-Ansatz}, 
\eqref{HW5D:EinsteinEqs:Ray} reduces to 
\Eq{
\pd_{\mu}\phi =\frac{9\pd_z h_1}{4h^2\pd_z\phi}\pd_{\mu}h_0.
}
Under the condition $\pd_z h_1\not=0$, this equation is equivalent to 
\Eq{
\phi= \Phi(h_0,h_1),\qquad
\Phi_0 \Phi_1=\frac{9}{4h^2},
\label{HW5D:phi}
}
where $\Phi_0 \equiv \pd_{h_0} \Phi$ and  $\Phi_1 \equiv \pd_{h_1} \Phi$. 

With the help of these relations, \eqref{HW5D:EinsteinEqs:RabOD}
can be written 
\Eqr{
&& R_{\mu\nu}(\Xsp) -\frac{1}{4} g_{\mu\nu}(\Xsp)R(\Xsp)
   +\inpare{\frac{9}{8h^2}-\frac{1}{2}\Phi_0^2}
    \insbra{ D_\mu h_0 D_\nu h_0 -\frac{1}{4} (Dh_0)^2 g_{\mu\nu}(\Xsp)}
\notag\\
&&\qquad
    -\frac{1}{h}\left [D_\mu D_\nu h_0
       -\frac{1}{4}\triangle_{\rm X} h_0 g_{\mu\nu}(\Xsp)\right]=0.
\label{HW5D:EinsteinEqs:RabOD-a}
}
Differentiating this equation by $y$, we get 
\Eq{
\inpare{\frac{9}{4h}+ h^2\Phi_0 \Phi_{01}}
   \insbra{ D_\mu h_0 D_\nu h_0 -\frac{1}{4} (Dh_0)^2 g_{\mu\nu}(\Xsp)}
 = D_\mu D_\nu h_0 -\frac{1}{4}\triangle_{\rm X} h_0 g_{\mu\nu}(\Xsp),
  \label{HW5D:EinsteinEqs:RabOD-1}
}
where $\Phi_{01}\equiv \pd_{h_0}\pd_{h_1}\Phi$. 
The factor in the square bracket on the left-hand side 
of this equation does not vanish under the condition 
$\pd_{\mu}h_0\ne 0$ because of the regularity of 
$g_{\mu\nu}$ as a matrix, and the right-hand side 
does not depend on $z$. Hence, the first factor on 
the left-hand side should be independent of $z$: 
\Eq{
0=\pd_{h_1} \inpare{\frac{9}{4h}+ h^2\Phi_0 \Phi_{01}}
  =\frac{9}{4}\pd_{h_0}\pd_{h_1} \ln (h \Phi_1).
}
Solving this with respect to $\Phi_1$ and using 
\eqref{HW5D:phi}, we obtain
\Eq{
\Phi_0=\frac{9}{4h a(h_0)b(h_1)},\quad
\Phi_1= \frac{a(h_0)b(h_1)}{h}.
    \label{HW5D:Phi0Phi1-2}
}
The consistency of these equations, 
$\pd_{h_1}\Phi_0=\pd_{h_0}\Phi_1$,  leads to
\Eq{
\frac{4}{9}\left[-a^2 + h a \left(\pd_{h_0}a\right)\right]
  +\frac{\pd_{h_1}b}{b^3} h + \frac{1}{b^2}=0.
   \label{HW5D:consistency}
}
Differentiating this equation by $h_0$ yields 
\Eq{
 \frac{4}{9}a \left(\pd_{h_0} a\right) 
 + h_0 \pd_{h_1}\pfrac{\pd_{h_1} b}{b^3} 
 -\frac{\pd_{h_1} b}{b^3}
   + h_1 \pd_{h_1}\pfrac{\pd_{h_1}b}{b^3}=0.
   \label{HW5D:consistency-2}
}
This equation implies that $a (\pd_{h_0} a)$ is 
a linear function of $h_0$. Hence, we get
\Eq{
a^2=p h_0^2 + 2 q h_0 + s,\qquad
\frac{1}{b^2}=\frac{4}{9}\inpare{ p h_1^2-2q h_1 + s},
}
where $p, q$ and $s$ are constant parameters. 
Inserting these expressions into \eqref{HW5D:Phi0Phi1-2}, we obtain 
\Eq{
\Phi_0=\pm \frac{3}{2h}\sqrt{\frac{ph_1^2-2q h_1 + s}{ph_0^2+2qh_0+s}},
\quad
\Phi_1=\pm \frac{3}{2h}\sqrt{\frac{ph_0^2+2qh_0+s}{ph_1^2-2q h_1 + s}}.
    \label{HW5D:Phi0Phi1-3}
}
This can be integrated to yield
\Eq{
e^{\mp 2\phi/3}
 =\frac{c}{h}\insbra{- p h_0 h_1 + q (h_0-h_1)+ s
    +\sqrt{ph_0^2+2qh_0+s}\sqrt{ph_1^2-2q h_1 + s}},
  \label{HW5D:scalarfield}
}
where $c$ is a constant. 

Using this expression for $\phi$, 
\eqref{HW5D:EinsteinEqs:RabOD-a} can be rewritten as
\Eqr{
&& h \insbra{R_{\mu\nu}(\Xsp)-\frac{1}{4}g_{\mu\nu}(\Xsp) R(\Xsp)}
    -D_\mu D_\nu h_0 + \frac{1}{4} \triangle_{\rm X}h_0 g_{\mu\nu}(\Xsp)
    \notag\\
&&\qquad
   +\frac{9}{8}\frac{p(h_0-h_1)+2q}{ph_0^2+2qh_0+s}
    \insbra{ D_\mu h_0 D_\nu h_0 -\frac{1}{4} (Dh_0)^2 g_{\mu\nu}(\Xsp)}
  =0.
   \label{HW5D:EinsteinEqs:RabOD-2}
}
Note that the left-hand side of this equation depends on 
$h_1$ linearly. Thus, this equation can be decomposed 
into two equations
\Eqrsubl{HW5D:EinsteinEqs:RabOD-3}{
&& \!\!\!\!\!\!\!\!\!\!\!\!\!\!\!\!\!\!\!\!\!\!\!
   R_{\mu\nu}(\Xsp)-\frac{1}{4}g_{\mu\nu}(\Xsp) R(\Xsp)
  =\frac{9}{8}\frac{p}{ph_0^2+2qh_0+s}
    \insbra{ D_\mu h_0 D_\nu h_0 -\frac{1}{4} (Dh_0)^2 g_{\mu\nu}(\Xsp)},\\
&& \!\!\!\!\!\!\!\!\!\!\!\!\!\!\!\! \!\!\!\!\!\!\! 
   D_\mu D_\nu h_0 - \frac{1}{4} \triangle_{\rm X} h_0 g_{\mu\nu}(\Xsp)
  =\frac{9}{4}\frac{ph_0+q}{ph_0^2+2qh_0+s}
    \insbra{ D_\mu h_0 D_\nu h_0 -\frac{1}{4} (Dh_0)^2 g_{\mu\nu}(\Xsp)}.
}
Multiplying the second of these by $D^\nu h_0$, we obtain
\Eq{
D_\mu (Dh_0)^2 -\frac{1}{2} (\triangle_{\rm X} h_0) D_\mu h_0
  =\frac{27}{16} \frac{ph_0+q}{ph_0^2+2qh_0+s} (Dh_0)^2 D_\mu h_0.
}
{}From this, we find that if $h_0$ satisfies $(Dh_0)^2=0$, 
then $\triangle_{\rm X}h_0=0$ holds. On the other hand, 
if $(Dh_0)^2\ne 0$, this equation can be deformed as 
\Eq{
D_\mu\insbra{\ln (Dh_0)^2-\frac{27}{32}\ln(p h_0^2+2qh_0+s)}
  =\frac{\triangle_{\rm X} h_0}{2 (Dh_0)^2} D_\mu h_0.
}
This equation implies that both $(Dh_0)^2$ and 
$\triangle_{\rm X}h_0$ depend on $x^\mu$ only through 
$h_0$, i.e., can be regarded as functions of $h_0$. 

Next, we analyse \eqref{HW5D:EinsteinEqs:RabTr} 
and \eqref{HW5D:EinsteinEqs:Ryy}, which can be now written  
\Eqrsubl{HW5D:tmpEqs}{
&& \!\!\!\!\!\!\!\!\!\!\!\!\!\!\!\!\!\!\!\!\!
R(\Xsp)-\frac{2}{h}\triangle_{\rm X} h_0 
   + \inpare{\frac{9}{8h^2}-\frac{\Phi_0^2}{2}}
   (Dh_0)^2 -\frac{\pd_z^2 h_1}{h^{3/2}}
    +\frac{(\pd_z h_1)^2}{2h^{5/2}}
   =\frac{4m^2}{3} e^{2\phi}h^{1/2},
 \label{HW5D:tmpEq:1}\\
&& \!\!\!\!\!\!\!\!\!\!\!\!\!\!\!\!\!\!\!\!\!
 -\frac{\triangle_{\rm X} h_0}{2h}-\frac{\pd_z^2 h_1}{h^{3/2}}
   +\frac{(\pd_z h_1)^2}{h^{1/2}}\inpare{\frac{5}{4h^2}-\frac{\Phi_1^2}{2}}
    =\frac{m^2}{3} h^{1/2} e^{2\phi}.
 \label{HW5D:tmpEq:2}
}
Here, note that the first equation and the above 
argument imply that $R(\Xsp)$ can be regarded as 
a function of $h_0$. Further, by eliminating 
$e^\phi$ from these equations, we obtain
\Eq{
h^{3/2} A + h^{1/2}B
  + 3\pd_z^2h_1 
  + \frac{9}{2}(\pd_z h_1)^2
    \left[\frac{p(h_0-h_1)+2q}{ph_1^2-2qh_1+s}\right]=0,
\label{HW5D:tmpEq:3}
}
where $A$ and $B$ are defined by 
\Eq{
A=R(\Xsp) -\frac{9p}{8} \frac{(Dh_0)^2}{ph_0^2+2qh_0+s},\quad
B=\frac{9 (Dh_0)^2}{4}\frac{ph_0+q}{ph_0^2+2q h_0+s}.
\label{HW5D:ABEq}}
Note that $A$ and $B$ can be regarded as functions of 
$h_0$ from the above arguments. 

Differentiating \eqref{HW5D:tmpEq:3} by $y$ twice, we get 
\Eq{
-B+ h(3A+ 4\pd_{h_0}B)+h^2(6 \pd_{h0} A+ 4 \pd_{h_0}^2B)
  + h^3 \pd_{h_0}^2A=0.
}
Since the left-hand side of this equation is a 
polynomial of $h_1$, the coefficients of powers of 
$h$ should vanish separately. This requires $A=B=0$. 
Hence, \eqref{HW5D:tmpEq:3} is equivalent to 
\Eq{
 p=0,\quad 
 q (Dh_0)^2=0,\quad
 R(\Xsp)=0,\quad
 \pd_z^2 h_1 + \frac{3q (\pd_z h_1)^2}{s-2q h_1}=0.
\label{HW5D:h1Eqs}
}

This equation implies that $q=0$ if $(Dh_0)^2\not=0$. 
On the other hand, in the case of $(Dh_0)^2=0$, 
which requires $\triangle_{\rm X} h_0=0$, 
\eqref{HW5D:tmpEq:1} reduces to 
\Eq{
-h \pd_z^2 h_1 + \frac{1}{2} (\pd_z h_1)^2= \frac{4m^2}{3} h^3 e^{2\phi}.
}
The left-hand side of this equation is linear in $h_0$. 
Hence, taking account of \eqref{HW5D:scalarfield}, 
we find that there exists a solution for $h_1$ only 
when $q=0$ also in the case of $(Dh_0)^2=0$.  

Thus, we can assume that $p=q=R(\Xsp)=0$. Then, 
\eqref{HW5D:scalarfield} and the last equation of 
\eqref{HW5D:h1Eqs} reduce to 
\Eq{
 e^{\mp 2\phi/3}=\frac{2cs}{h},\qquad
 \pd_z h_1=k. 
}
Inserting these expressions into \eqref{HW5D:tmpEq:2}, 
we obtain
\Eq{
 e^{2\phi/3}=\frac{2cs}{h},\quad
 \frac{k^2}{(2cs)^3}= \frac{8m^2}{3},\quad
 \triangle_{\rm X} h_0=0.
 \label{HW5D:hscalar-condition}
}
Hence, \eqref{HW5D:EinsteinEqs:RabOD-3} reduces to 
$R_{\mu\nu}(\Xsp)=0$ and $D_\mu D_\nu h_0=0$. 

To summarise, under the conditions \eqref{HW5D:ap-solution} 
and \eqref{HW5D:ap-Ansatz}, the most general solution 
of the field equations \eqref{HW5D:FEqs} is given by 
\Eq{
R_{\mu\nu}(\Xsp)=0,\qquad
h(x, z)=h_0(x)+ k z, \qquad e^{2\phi}=l^2h^{-3},
    \label{HW5D:apHWsolution}
}
where $l$ is a constant, and $h_0$ and $k$ satisfy 
the conditions  
\Eq{
D_\mu D_\nu h_0=0,\qquad k^2=\frac{8}{3}m^2 l^2.
}
Here, we can set $l=1$ by redefining $k$, $z$ and 
$ds^2(\Xsp)$. Further, a solution with $(D h_0)^2\not=0$ 
exists only when $\Xsp$ is locally flat, and in that 
case, $h_0$ can be written as a linear combination of 
the Minkowski coordinates \cite{Kodama:2005fz}.  
This solution with Minkowskian $\Xsp$ is identical to 
the solution found by Chen et al. \cite{Chen:2005jp}.


\end{document}